\documentclass[12pt,a4paper,twoside]{article}
\usepackage{amsmath}
\usepackage{amssymb}
\usepackage{amsthm}
\newcommand*{\anno}[1]{\oldstylenums{#1}} 
\makeatletter
\newif\if@openbib       
\@openbibfalse          
\def\bibstyle@aa{\bibpunct{(}{)}{;}{a}{}{,}}
\def\bibstyle@pass{\bibpunct{(}{)}{;}{a}{,}{,}}
\def\bibstyle@anngeo{\bibpunct{(}{)}{;}{a}{,}{,}}
\def\bibstyle@agsm{\bibpunct{(}{)}{,}{a}{}{,}\gdef\harvardand{\&}}
\def\bibstyle@kluwer{\bibpunct{(}{)}{,}{a}{}{,}\gdef\harvardand{\&}}
\def\bibstyle@dcu{\bibpunct{(}{)}{;}{a}{;}{,}\gdef\harvardand{and}}
\def\bibstyle@agu{\bibpunct{[}{]}{;}{a}{,}{,}}
\def\bibstyle@nlinproc{\bibpunct{(}{)}{;}{a}{,}{,}}
\gdef\@oldstyle#1{#1}
\def\bibstyle@pmbib{\bibpunct{(}{)}{;}{a}{}{,}%
             \gdef\@oldstyle##1{\oldstylenums{##1}}}
\def\bibstyle@oldapa{\bibpunct{(}{)}{;}{a}{}{,}%
             \gdef\@oldstyle##1{\oldstylenums{##1}}}
\def\bibstyle@oldapanp{\bibpunct{(}{)}{;}{a}{}{,}%
             \gdef\@oldstyle##1{\oldstylenums{##1}}}
\def\bibstyle@oldapasc{\bibpunct{(}{)}{;}{a}{}{,}%
             \gdef\@oldstyle##1{\oldstylenums{##1}}}
\def\bibstyle@oldapascnp{\bibpunct{(}{)}{;}{a}{}{,}%
             \gdef\@oldstyle##1{\oldstylenums{##1}}}
\def\bibstyle@RoySoc{\bibpunct{(}{)}{;}{a}{}{,}%
             \gdef\@oldstyle##1{\oldstylenums{##1}}}
\def\bibstyle@RoySocNu{\bibpunct{(}{)}{;}{a}{}{,}}
\def\bibstyle@cospar{\bibpunct{/}{/}{,}{n}{}{}%
     \gdef\@biblabel##1{##1.}}
\def\bibstyle@esa{\bibpunct{(}{)}{,}{n}{}{}%
     \gdef\@biblabel##1{##1.\hspace{1em}}%
     \gdef\@cite##1##2##3{\@citebegin Ref.~##1\if@tempswa,
          ##3\fi\@citeend}}
\def\bibstyle@nature{\bibpunct{}{}{,}{n}{}{}%
     \gdef\@biblabel##1{##1.}%
     \gdef\@cite##1##2##3{\unskip\mbox{$^{##1}$}}}
\def\bibstyle@plain{\bibpunct{[}{]}{,}{n}{}{}}
\let\bibstyle@alpha=\bibstyle@plain
\let\bibstyle@abbrv=\bibstyle@plain
\let\bibstyle@unsrt=\bibstyle@plain
\InputIfFileExists{pmbib.cfg}
       {\typeout{Local config file pmbib.cfg used}}{}
\def\@bibtocline{}
\def\@smallbibsize{}
\def\@cite#1#2#3{\if@tempswa\@citebegin\if#2\@empty\else#2 \fi
        #1\if#3\@empty\else, #3\fi\@citeend\else#1\fi}
\def\@citenum#1#2#3{\@citebegin\if@tempswa\if#2\@empty\else#2 \fi\fi
   #1\if#3\@empty\else, #3\fi\@citeend}
\def\@citexnum[#1][#2]#3{\if@filesw\immediate\write
      \@auxout{\string\citation{#3}}\fi
  \let\@citea\@empty
  \@cite{\@for\@citeb:=#3\do
    {\@citea\def\@citea{\@citesep\penalty\@m\ }%
     \def\@tempa##1##2\@nil{\edef\@citeb{\if##1\space##2\else##1##2\fi}}%
     \expandafter\@tempa\@citeb\@nil
     \@ifundefined{b@\@citeb}{%
       {\reset@font\bfseries ?}\G@refundefinedtrue\@latex@warning
       {Citation `\@citeb' on page \thepage \space undefined}}%
     \hbox{\csname b@\@citeb\endcsname}}}{#1}{#2}}
\def\@citex[#1][#2]#3{\let\@citea\@empty
  \@cite{\let\@citenm\@empty
    \@for\@citeb:=#3\do
    {\edef\@citeb{\expandafter\@iden\@citeb}%
     \if@filesw\immediate\write\@auxout{\string\citation{\@citeb}}\fi
     \@ifundefined{b@\@citeb}{\@citea%
       {\reset@font\bfseries ?}\G@refundefinedtrue\@latex@warning
       {Citation `\@citeb' on page \thepage \space undefined}}%
     {\let\@citemm=\@citenm
     \@cite@parse{\@citeb}%
     \if@tempswa
       \ifx\@citemm\@citenm\@yrsep\else\@citea{\@citenm}\@auyrsep\fi
       \ \@citedt \def\@citea{\@citesep\ }%
     \else
       \ifx\@citemm\@citenm, \@citedt\else\@citea{\@citenm}~
           \@citebegin\@citedt\fi
       \def\@citea{\@citeend\@citesep\ }%
     \fi}}\if@tempswa\else\@citeend\fi}{#1}{#2}}
\def\@biblabel#1{\hfill}
\def\@biblabelnum#1{[#1]}
\def\@bibsetnum#1{\settowidth\labelwidth{\@biblabel{#1}}%
   \leftmargin\labelwidth \advance\leftmargin\labelsep
   \if@openbib
     \advance\leftmargin\bibindent
     \itemindent -\bibindent
     \listparindent \itemindent
     \parsep \z@
   \fi
}
\def\@bibsetup#1{\leftmargin=1em\itemindent=-\leftmargin}
\def\@auyrsep{,} \def\@yrsep{,}
\def\bibstyle#1{\@ifundefined{bibstyle@#1}{\relax}
     {\csname bibstyle@#1\endcsname}}
\AtBeginDocument{\global\let\bibstyle=\@gobble}
\def\bibpunct#1#2#3#4#5#6{\gdef\@citebegin{#1}\gdef\@citeend{#2}\gdef
   \@citesep{#3}\ifx #4n\global\let\@bibsetup=\@bibsetnum
   \global\let\@citex=\@citexnum
   \global\let\@biblabel=\@biblabelnum
   \global\let\@cite=\@citenum\fi
   \gdef\@auyrsep{#5}\gdef\@yrsep{#6}}
\newif\ifNAT@full\NAT@fullfalse
\def\cite{\@ifstar{\NAT@fulltrue\@citee}{\NAT@fullfalse\@citee}}
\def\@citee{\@ifnextchar [{\@tempswatrue\@citex@}{\@tempswafalse
    \@citex@[]}}
\def\@citex@[#1]{\@ifnextchar [{\@citex[#1]}{\@citex[][#1]}}
\newcommand{\citeauthor}[1]{\if@filesw\immediate\write
     \@auxout{\string\citation{#1}}\fi
\ifx\@citex\@citexnum
       {\reset@font\bfseries(author?)}\PackageWarning{pmbib}
       {Cannot use \protect\citeauthor
         \MessageBreak
        with numerical citations}\else
     \@ifundefined{b@#1}{%
       {\reset@font\bfseries ?}\G@refundefinedtrue\@latex@warning
       {Citation `#1' on page \thepage \space undefined}}%
       {\@cite@parse{#1}{\@citenm}}\fi}
\newcommand{\citeyear}[1]{\if@filesw\immediate\write
   \@auxout{\string\citation{#1}}\fi
\ifx\@citex\@citexnum
       {\reset@font\bfseries(year?)}\PackageWarning{pmbib}
       {Cannot use \protect\citeyear
         \MessageBreak
        with numerical citations}\else
     \@ifundefined{b@#1}{%
       {\reset@font\bfseries ?}\G@refundefinedtrue\@latex@warning
       {Citation `#1' on page \thepage \space undefined}}%
       {\@cite@parse{#1}\@citedt}\fi}
\newcommand{\citefullauthor}[1]{\if@filesw\immediate\write
     \@auxout{\string\citation{#1}}\fi
\ifx\@citex\@citexnum
       {\reset@font\bfseries(author?)}\PackageWarning{pmbib}
       {Cannot use \protect\citeauthor
         \MessageBreak
        with numerical citations}\else
     \@ifundefined{b@#1}{%
       {\reset@font\bfseries ?}\G@refundefinedtrue\@latex@warning
       {Citation `#1' on page \thepage \space undefined}}%
       {\@cite@parse{#1}{\@citefull}}\fi}
\def\@cite@parse#1{{%
     \let\protect=\@unexpandable@protect
     \xdef\@tempa{\csname b@#1\endcsname\relax}}%
     \expandafter\@citez\@tempa()()\@nil
     \ifNAT@full\let\@citenm\@citefull\fi}
\def\@citez#1(#2)#3()#4\@nil{\gdef\@citenm{#1}\gdef\@citedt{#2}%
  \ifx#3\relax\gdef\@citefull{#1}\else\gdef\@citefull{#3}\fi
  \if!#2!\expandafter\@citeapalk#1\@nil\fi}
\def\@citeapalk#1, #2\@nil{\gdef\@citenm{#1}\gdef\@citedt{\@oldstyle{#2}}%
   \gdef\@citefull{#1}}
\newcommand{\citeauthoryear}[3]{\ifx#3\relax #1(#2)#1\else #2(#3)#1\fi}
\newcommand{\citestarts}{\@citebegin}
\newcommand{\citeends}{\@citeend}

\def\harvarditem{\@ifnextchar[{\@harvarditem}{\@harvarditem[\@empty]}}
\def\@harvarditem[#1]#2#3#4{\if!#1!\bibitem[#2(#3)]{#4}\else
  \bibitem[#1(#3)#2]{#4}\fi }
\newcommand{\harvardleft}{\@citebegin}
\newcommand{\harvardright}{\@citeend}
\newcommand{\harvardyearleft}{\@citebegin}
\newcommand{\harvardyearright}{\@citeend}
\AtBeginDocument{\providecommand{\harvardand}{and}}

\@ifundefined{chapter}{\def\bibsection{\section*{\refname
   \@mkboth{\uppercase{\refname}}{\uppercase{\refname}}}}}{\def
   \bibsection{\chapter*{\bibname
   \@mkboth{\uppercase{\bibname}}{\uppercase{\bibname}}}}}
\renewenvironment{thebibliography}[1]
 {\bibfont\bibsection\parindent \z@\list
   {\@biblabel{\arabic{enumiv}}}{\@bibsetup{#1}%
    \usecounter{enumiv}\let\p@enumiv\@empty
    }%
    \if@openbib
      \renewcommand\newblock{\par}
    \else
      \renewcommand\newblock{\hskip .11em \@plus.33em \@minus.07em}%
    \fi
    \sloppy\clubpenalty4000\widowpenalty4000
    \sfcode`\.=1000\relax\@smallbibsize\@bibtocline}
  {\def\@noitemerr{%
  \@latex@warning
     {Empty `thebibliography' environment}}%
  \endlist\vskip-\lastskip}
\let\bibfont=\relax
\def\@mn{\oldstylenums}
\def\today{\ifcase\day\or
  \@mn{1}st\or \@mn{2}nd\or \@mn{3}rd\or \@mn{4}th\or \@mn{5}th\or
  \@mn{6}th\or \@mn{7}th\or \@mn{8}th\or \@mn{9}th\or \@mn{10}th\or
  \@mn{11}th\or \@mn{12}th\or \@mn{13}th\or \@mn{14}th\or \@mn{15}th\or
  \@mn{16}th\or \@mn{17}th\or \@mn{18}th\or \@mn{19}th\or \@mn{20}th\or
  \@mn{21}st\or \@mn{22}nd\or \@mn{23}rd\or \@mn{24}th\or \@mn{25}th\or
  \@mn{26}th\or \@mn{27}th\or \@mn{28}th\or \@mn{29}th\or \@mn{30}th\or
  \@mn{31}st\fi~\ifcase\month\or
  January\or February\or March\or April\or May\or June\or
  July\or August\or September\or October\or November\or December\fi
  \space \oldstylenums{\number\year}}
\def\peqref#1{\textup{\tagform@{\ref{#1}$'$}}}
\def\ppeqref#1{\textup{\tagform@{\ref{#1}$''$}}}

\renewenvironment{subequations}{%
  \refstepcounter{equation}%
  \begingroup 
  \let\protect\@nx
  \edef\@tempa{\def\@nx\theparentequation{\theequation}}%
  \@xp\endgroup\@tempa
  \setcounter{parentequation}{\value{equation}}%
  \setcounter{equation}{0}%
  \def\theequation{\theparentequation\oldstylenums{\alph{equation}}}%
  \ignorespaces
}{%
  \setcounter{equation}{\value{parentequation}}%
  \global\@ignoretrue
}
\newcommand{\leftsideset}[2]{%
  \@mathmeasure\z@\displaystyle{#2}%
  \global\setbox\@ne\vbox to\ht\z@{}\dp\@ne\dp\z@
  \setbox\tw@\box\@ne
  \@mathmeasure4\displaystyle{\copy\tw@#1}%
  \@mathmeasure6\displaystyle{#2}%
  \dimen@-\wd6 \advance\dimen@\wd4 \advance\dimen@\wd\z@
  \hbox to\dimen@{}\mathord{\kern-\dimen@\box4\box6}{}%
}
\makeatother
\catcode`\ˆ=\active\defˆ{\`a}
\catcode`\=\active\def{\`e}
\catcode`\Ž=\active\defŽ{\'e}
\catcode`\'=\active\def'{\'{\ii}}
\catcode`\˜=\active\def˜{\`o}
\catcode`\œ=\active\defœ{\'u}
\catcode`\é=\active\defé{\`E}
\catcode`\£=\active
\title{\itshape Metric-affine gravity\\[1ex]
       and\\[1ex]
       the Nester-Witten 2-form}
\author{Marco Godina\thanks{Dipartimento di Matematica, Universitˆ di 
        Torino, Via Carlo Alberto~10, 10123 Torino, Italy.}, 
        Paolo Matteucci\thanks{Faculty of Mathematical Studies, 
        University of Southampton, Highfield, Southampton SO17 1BJ, 
        England (UK).}\ \thanks{Corresponding author. \emph{E-mail address}:
        \texttt{p.matteucci@maths.soton.ac.uk}.}
        \ \& James A. Vickers$^\dagger$}

\newcommand{£}{\pounds\ida}
\DeclareMathOperator{\hor}{Hor}

\DeclareMathOperator{\inn}{\rfloor}
\let\Re=\relax
\DeclareMathOperator{\Re}{Re}
\newcommand{\thorn}{\hbox{\textsf{\setbox0=\hbox{l}\copy0\kern-\wd0 p}}}
\newcommand{\A}{{\textstyle\bigwedge}}

\newcommand{\ag}{\mathfrak{a}}
\newcommand{\Aut}{\mathrm{Aut}}

\newcommand{\bl}{\bar\lambda}

\newcommand{\C}{\mathbb{C}}
\newcommand{\cc}{\textsc{cc}}
\newcommand{\cd}{\nabla\ida}
\newcommand{\cf}{\protect\emph{cf.{}}}

\renewcommand{\d}{\mathrm{d}}
\newcommand{\de}{\partial}
\newcommand{\dH}{\d_{\mathrm{H}}}

\newcommand{\dV}{\d_{\mathrm{V}}}

\newcommand*{\emphbf}[1]{\emph{\textbf{#1}}}

\newcommand*{\ftimes}[1]{\underset{#1}{\times}}

\newcommand{\GL}{\mathrm{GL}}
\renewcommand{\i}{\mathrm{i}}

\newcommand*{\ida}[2]{\ifx#1^{}^{#2}
                      \else\ifx#1_{}_{\!#2}
                      \else\errmessage{Sub/Superscript token missing}\fi
                      \fi}

\renewcommand{\l}{\lambda}
\renewcommand{\L}{\mathcal{L}}

\newcommand{\LD}{\L_{\mathrm{D}}}
\newcommand{\LL}{\mathbb{L}}
\newcommand{\liestar}[1]{\leftsideset{^*}{\!#1}}

\newcommand{\nw}{Nester-Witten}

\newcommand{\pc}{PoincarŽ-Cartan}
\newcommand{\pf}{\protect{\S}}
\newcommand{\pr}{\mathrm{pr}}
\newcommand{\R}{\mathbb{R}}

\renewcommand{\sl}{\mathfrak{sl}}
\newcommand{\SL}{\mathrm{SL}}
\newcommand{\so}{\mathfrak{so}}
\newcommand{\SO}{\mathrm{SO}}
\newcommand{\spin}{\mathfrak{spin}}
\newcommand{\Spin}{\mathrm{Spin}}

\newcommand{\VXi}{\check\Xi}
\newcommand{\VxiCW}{\check\xi_{\mathrm{W}}^{\C}{}}
\newcommand{\VxiK}{\check\xi_{\mathrm{K}}{}}
\newcommand{\VxiW}{\check\xi_{\mathrm{W}}{}}

\newcommand{\X}{\mathfrak{X}}

\newcommand{\xiCW}{\xi_{\mathrm{W}}^{\C}{}}
\newcommand{\xiK}{\xi_{\mathrm{K}}{}}
\newcommand{\xiW}{\xi_{\mathrm{W}}{}}
\newcommand{\Z}{\mathbb{Z}}

\renewcommand{\[}{\left[}

\newtheorem{prop}{Proposition}[section]

\theoremstyle{definition}
\newtheorem{rem}[prop]{Remark}
\numberwithin{equation}{section}

\begin{document}
\hyphenation{Frau-en-die-ner Gia-chet-ta Hil-bert rai-fear-taigh}

\maketitle
\begin{abstract}
    In this paper we redefine the well-known metric-affine Hilbert Lagrangian 
    in terms of a spin-connection and a spin-tetrad. On applying the \pc\
    method and using the geometry of gauge-natural bundles, a global
    gravitational superpotential is derived. On specializing to the case of the
    Kosmann lift, we recover the result originally found by \cite{kijowski78} for the
    metric (natural) Hilbert Lagrangian. On choosing a different, suitable
    lift, we can also recover the \nw\ 2-form, which plays an important role in
    the energy positivity proof and in many quasi-local definitions of mass.
\end{abstract}

\pagestyle{myheadings}
\markboth{\itshape Marco Godina, Paolo Matteucci \& James A. Vickers}%
         {\itshape Metric-affine gravity and the Nester-Witten 2-form}
\thispagestyle{plain}

\section*{Introduction}

Conserved quantities have always represented an intriguing issue in general 
relativity, as was pointed out by \cite{penrose82} in a very famous paper. The jet bundle
formalism provides an adequate framework for Lagrangian  field theories and the \pc\ method
enables one to  associate with each of them globally conserved charges  \cite[\cf{, e.g.,}
][]{trautman67,krupka71,gms}. In particular, for  first order theories these charges are
uniquely defined, and in the  second order case, although uniqueness is lost, still there is a
unique \emph{canonical} choice.

Natural Lagrangian field theories have been known for a long time, 
Einstein's general relativity being one of them. Many physical 
theories, though, such as Yang-Mills and Dirac theories, are \emph{non-%
natural}, i.e.\ the ``configuration bundle'', which is nothing but the space 
of the dependent variables or ``fields'', is not a natural bundle.
Roughly speaking, \emph{natural bundles} (such as the tangent or the cotangent bundle)
form a particular class of fibre bundles, where, once a coordinate change on the base
manifold is given, the corresponding fibred coordinate change is known.
More technically, natural bundles can be regarded as fibre bundles associated with
higher order frame bundles on manifolds \cite[\cf\ ][]{kms}.

If we aim at considering the coupling of a natural theory, 
such as general relativity, with a non-natural one, we are \emph{sometimes} 
forced to ``redefine'' our field variables in order to make the coupling
physically meaningful. In particular, if we want to describe the interaction  and
feedback between gravity and spinor fields, \emph{spin-tetrads}, and  not tetrads,
are the appropriate objects to be considered \cite[\cf\ ][]{fffg98,gmff}.
\emph{Gauge-natural bundles} provide a suitable geometrical
framework for such objects.  These bundles are fibre bundles
associated with ``abstract'' principal bundles with arbitrary structure group
\cite[\cf\ ][]{kms}.

The superpotential associated with the standard Hilbert Lagrangian for 
general relativity, or the ``Hilbert superpotential'', was first given by
\cite{kijowski78} using ideas developed by \citeauthor{kijowski73} himself
(\citeyear{kijowski73}) and \cite{ks75,ks76}. It was also derived  explicitly using 
the \pc\ method by \cite{kt79}, in a Hamiltonian (multisymplectic) framework, and \cite{ffr86},
in the Lagrangian context\footnote{There exists an extensive literature on both the
multisymplectic formulation of field theories and its Lagrangian counterpart. The interested
reader is referred to \cite{gimmsy1}, on page~15 and 25--6, respectively}. In \cite{ffm94} the
authors were able to reformulate the  previous result of two of them in terms of tetrads. But,
again, their theory was still natural,  and this meant there was no real advantage of such a
reformulation.

Recently, two parallel papers \cite[]{fffg98,gmff} addressed the problem of
re-expressing the above results in terms of spin-tetrads and coupling \emph{true}
general relativity with Fermionic matter, but their findings implicitly relied on
a \pc\ form associated with a particular (``quasi-natural'') lift of vector fields
onto the  bundle of orthonormal frames, the ``Kosmann lift'' 
\cite[\cf\ ][]{fffg96}.

In this paper we redefine the \emph{metric-affine} Hilbert Lagrangian in terms 
of a spin-connection and a spin-tetrad. The ensuing superpotential is 
genuinely ``general'', in the sense that it is derived in a \emph{completely}
gauge-natural context, and also allows for the presence of torsion.

Such a reformulation enables us not only to single out the aforementioned link
between the Hilbert superpotential and the Kosmann lift, but also to associate the
well-known \nw\ 2-form with another particular lift, thereby providing us with a
clear-cut geometric  interpretation of a rather famous but somewhat obscure
integrand in  general relativity. 
This lift turns out to be essentially the \emph{dual} of the Kosmann
lift. For a different characterization of
the \nw\ 2-form see, e.g., the detailed analysis by \cite{dubois-madore87}.

The structure of the paper is as follows: in \S\ref{sec:pc} we recall the main
ingredients of the \pc\ method, in \S\ref{sec:gf} we set up the geometric framework
of our theory,  and in \S\ref{sec:mag} we derive our main results.

Finally, in \S\ref{sec:fog} we present a first order covariant Lagrangian for
general relativity and derive the relevant superpotential.

\section{\pc\ method}
\label{sec:pc}

It is well-know that to each \emph{first order} Lagrangian there corresponds a
\emph{unique} global 
\pc\ form. Let $M$ be an (orientable, Hausdorff, paracompact, smooth)
$m$-dimensional manifold and
\begin{equation*}
    \left\{
\begin{aligned}
	{}&\L \colon J^1\!B\to\A^{m} T^{*}\!M  \\
    {}&\L \colon j^1y\mapsto 
       \L(j^1y) \equiv L(x^\l,y^\ag,y^\ag\ida_\mu)\,\d s
\end{aligned}
\right.
\end{equation*}
a first order Lagrangian defined on the first order jet 
prolongation $J^{1}\!B$ of a gauge-natural bundle~$B$ over~$M$ 
\cite[\cf\ ][\pf51]{kms}, $(x^\l,y^\ag)$ being coordinates on~$B$ and $\d
s\equiv \d x^{0}\wedge \d x^{1}\wedge \dots\wedge \d x^{m-1}$ 
the natural volume element on~$M$.  Define its \emphbf{momenta} as
\begin{equation*}
f_{\ag}\ida^{\mu}:=\frac{\de L}{\de y^{\ag}\ida_{\mu}}.
\end{equation*}
The \emphbf{\pc\ form} associated with~$\L$ is then given by
\begin{equation*}
\Theta(\L):= \L+f_{\ag}\ida^{\mu}\,\dV y^{\ag}\wedge\d s_{\mu},
\end{equation*}
where $\dV$ is the vertical differential (notably, 
$\dV y^{\ag}=\d y^{\ag}-y^{\ag}\ida_{\mu}\,\d x^\mu$: \cf~%
\citeauthor{gms}\/\ \citeyear{gms}) and we set 
$\d s_{\mu}:=\de_{\mu}\inn\d s$, `$\inn$' denoting the inner product.

The knowledge of the \pc\  form enables us to calculate the
so-called \emphbf{Noether current} of the Lagrangian in question. 
Indeed, if one has a one-parameter subgroup of 
automorphisms of~$B$ generated by a projectable vector field~$\Xi$ (with projection~$\xi$
onto~$M$), the Noether current associated with~$\L$ along the  vector field~$\Xi$ is given by
\begin{align*}
    E(\L,\Xi)&:=-\hor[J^1\Xi\inn\Theta(\L)]  \\
             &\hphantom{:}= -\xi\inn\L + f_{\ag}\ida^{\mu}£_{\Xi}y^{\ag}
                                         \,\d s_{\mu},
\end{align*}
where $\hor$ denotes the horizontal projection \cite[\cf\ ][\pf3.1]{gms}, 
$J^1\Xi$ is the first order jet prolongation of~$\Xi$, 
and the well-known relation
\begin{equation*}
    J^{1}\Xi\inn\dV y^{\ag} = -£_{\Xi}y^{\ag}
    \label{eq:Lie-dV}
\end{equation*}
between vertical differential and (generalized) Lie derivative 
is used in obtaining the second equation \cite[\cf\ ][\pf47]{kms}.

\section{Geometric framework}
\label{sec:gf}

Let $M$ be an orientable, Hausdorff,
paracompact, smooth, 4-dimensional manifold.
Suppose $M$ admits Lorentzian metrics of signature $-2$, i.e.\
assume that $M$ satisfies the topological requirements which ensure 
the existence on it of  Lorentzian structures [$\SO(1,3)^e$-reductions]. 
Let $\LL(M)$ be the (principal) bundle of linear frames over~$M$ with 
structure group $\GL(4,\R)$.

Assume now that $M$ admits a \emphbf{free spin structure}
$(\Sigma,\tilde\Lambda)$,  i.e.\ the existence of at least one principal (fibre)
bundle $\Sigma$  over $M$ with structure group $\Spin(1,3)^e\cong\SL(2,\C)$,
called the \emphbf{spin structure bundle}, and at least one strong (i.e.\ covering
the identity map) equivariant morphism $\tilde\Lambda\colon\Sigma\to \LL(M)$
\cite[]{gmff}. We call the bundle map $\tilde\Lambda$ a \emphbf{spin-frame} on
$\Sigma$.

This definition of a spin structure induces metrics on $M$.
Indeed, given a spin-frame $\tilde\Lambda\colon\Sigma\to \LL(M)$,
we can define a metric via the reduced subbundle $\SO(M,g)\equiv
\tilde\Lambda(\Sigma)$ of $\LL(M)$. In other words, the \emph{dynamic} metric $g\equiv
g_{\tilde\Lambda}$ is defined to be the metric such that
frames in $\tilde\Lambda(\Sigma)\subset \LL(M)$ are $g$-orthonormal frames.
It is important to stress that in our picture the metric~$g$
is built up \emph{a posteriori}, after a spin-frame has been determined by the 
field equations in a way which is compatible with the (free) spin structure
one has used to define spinors.

Now let $\Lambda$ be the epimorphism which exhibits $\Spin(1,3)^e$  as a twofold 
covering of $\SO(1,3)^e$ and consider the following left action of the group 
$\GL(4,\R)\times\Spin(1,3)^e$ on the manifold $\GL(4,\R)$
\begin{equation*}
\left\{
\begin{aligned}
\rho&\colon(\GL(4,\R)\times\Spin(1,3)^e)\times \GL(4,\R)\to \GL(4,\R)  \\
\rho&\colon((A^\mu\ida_\nu,S^{a}\ida_{b}),u^a\ida_\mu)
\mapsto u'{}^{a}\ida_{\mu}
        :=(\Lambda(S))^a\ida_b u^b\ida_\nu(A^{-1})^\nu\ida_\mu
\end{aligned}
\right.
\end{equation*}
together with the associated bundle $\Sigma_\rho:=W^{1,0}(\Sigma)
\times_{\rho}\GL(4,\R)$, where $W^{1,0}(\Sigma):=\LL(M)\ftimes{M}\Sigma$
denotes the principal prolongation of order~$(1,0)$ of the principal fibre
bundle~$\Sigma$ \cite[\cf\ ][\pf52.4]{kms}.
The bundle $W^{1,0}(\Sigma)$ is a principal fibre bundle with
structure group $\GL(4,\R)\times\Spin(1,3)^e$.
It turns out that $\Sigma_\rho$ is a fibre bundle associated with $W^{1,0}(\Sigma)$, 
i.e.\ a gauge-natural bundle of order~$(1,0)$. A section
of $\Sigma_\rho$ will be called a \emphbf{spin-tetrad}.

Recall now that a \textbf{(\emph{principal}) \emph{connection}} on a principal (fibre) 
bundle $P(M,G)$ may be regarded as a $G$-equivariant global 
section of the affine jet bundle $J^{1}\!P\to P$, where the $G$-action on $J^1\!P$
is induced by the first jet prolongation of the canonical
(right) action of~$G$ on~$P$ \cite[\cf\ ][\pf2.7]{gms}. 
Owing to $G$-equivariance there is a 1-1 correspondence between 
principal connections and global sections of the quotient bundle 
$J^{1}\!P/G\to M$.

More specifically, let $P=\Sigma$ and let $\spin(1,3)\cong\so(1,3)\cong\sl(2,\C)$ 
denote the Lie algebra of $\Spin(1,3)^e$. Consider then the following left action 
on the vector space $V_C :=(\R^{4})^{*}\otimes\so(1,3)$
\begin{equation*}
\left\{
\begin{aligned}
\lambda&\colon(\GL(4,\R)\times T^{1}_{4}\Spin(1,3)^e)\times V_C\to V_C  \\
\lambda&\colon((A^\mu\ida_\nu,S^{a}\ida_{b},S^{a}\ida_{b\mu}),
               u^{a}\ida_{b\mu})
         \mapsto u'{}^{a}\ida_{b\mu} 
         := (A^{-1})^{\nu}\ida_{\mu}
     [(\Lambda(S))^{a}\ida_{c}u^{c}\ida_{d\nu}(\Lambda(S^{-1}))^{d}\ida_{b}  \\
     &\qquad\qquad\qquad\qquad\qquad\qquad\qquad\qquad\qquad\qquad
     {}-(\Lambda(S))^{a}\ida_{c\nu}(\Lambda(S^{-1}))^{c}\ida_{b}]
\end{aligned}
\right.,
\end{equation*}
where $(\Lambda(S))^a\ida_{c\nu}$ are the components of
$j^1_0(\Lambda\circ S)$, an element of $T^1_4\SO(1,3)^e$,
and $S\colon\R^4 \to\Spin(1,3)^e$ is a local map defined around
the origin $0\in\R^4$. Hence define the associated bundle 
$C:=W^{1,1}(\Sigma)\times_{\lambda}V_C$, where
$W^{1,1}(\Sigma):=\LL(M)\ftimes{M} J^{1}\Sigma$ denotes the principal 
prolongation of order~$(1,1)$ of~$\Sigma$ 
\cite[\cf\ ][\pf52.4]{kms}. It turns out that $C$ 
is a gauge-natural bundle of order~$(1,1)$ isomorphic to $J^1(\Sigma/\Z_2)/\SO(1,3)^e$. 
A section of~$C$ will be called a \emphbf{spin-connection}.

\section{Metric-affine gravity}
\label{sec:mag}

Let $\theta^{a}\ida_{\mu}$ be a spin-tetrad and 
$\omega^{a}\ida_{b\mu}$ a spin-connection, as defined in the previous section. 
Set locally
\begin{align*}
	\theta^{a} &:= \theta^{a}\ida_{\mu}\,\d x^{\mu},  \\
      e_{a} &:= e_{a}\ida^{\mu}\de_{\mu},
\end{align*}
where $e_{a}\ida^{\mu}$ is implicitly defined via the relation
$\theta^{a}\ida_{\mu}e_{b}\ida^{\mu}=\delta^{a}\ida_{b}$, and
\begin{align*}
 \omega^{a}\ida_{b} &:= \omega^{a}\ida_{b\mu}\,\d x^{\mu},  \\
 \Omega^{a}\ida_{b} &:= \dH\omega^{a}\ida_{b}
                        + \omega^{a}\ida_{c}\wedge\omega^{c}\ida_{b},
\end{align*}
$\dH$ being the horizontal differential \cite[\cf\ ][\pf3.1]{gms}; 
$\omega^{a}\ida_{b}$ and $\Omega^{a}\ida_{b}$ are recognized to be 
the (horizontal) connection 1-form and curvature 2-form, respectively.

We can now ``redefine'' the (metric-affine) \emphbf{Hilbert Lagrangian} as
\begin{equation}
    \left\{
\begin{aligned}
	{}&\L \colon \Sigma_{\rho}\ftimes{M}
	       J^{1}C\to\A^{4} T^{*}\!M  \\
    {}&\L \colon (\theta^{a}\ida_{\mu},j^{1}\omega^{a}\ida_{b\mu})\mapsto
       \L (\theta^{a}\ida_{\mu},j^{1}\omega^{a}\ida_{b\mu})
       := -\tfrac{1}{2\kappa}\Omega_{ab}\wedge\Sigma^{ab}
\end{aligned}
\right.,
    \label{eq:Lec}
\end{equation} 
where $\kappa:=8\pi G/c^4$ and $\Sigma^{ab}:=\liestar{}(\theta^{a}\wedge\theta^{b})$. 
The equations of motion are obtained by varying $\L$ with respect to~$\theta^{c}$ 
and~$\omega_{ab}$:
\begin{subequations}
\begin{align}
	\tfrac{\delta\L}{\delta\theta^{c}} & \equiv 
	  \tfrac{1}{2\kappa}\Omega_{ab}\wedge\Sigma^{ab}\ida_{c} \equiv 
	  -\tfrac{1}{\kappa}G^{a}\ida_{c}\Sigma_{a} = 0,
	\label{eq:moto1}  \\
	\tfrac{\delta\L}{\delta\omega_{ab}} & \equiv
	  \tfrac{1}{2\kappa}\nabla\Sigma^{ab} = 0,
	\label{eq:moto2}
\end{align}
\end{subequations}
where $\Sigma^{ab}\ida_{c}:=e_{c}\inn\Sigma^{ab}$, $\Sigma_a:=
1/6\,e_{abcd}\,\theta^{b}\wedge\theta^{c}\wedge\theta^{d}$ and $\nabla$ denotes 
the \hbox{(gauge-)}\allowbreak covariant exterior derivative. We stress that the
condition
$\nabla\Sigma^{ab}=0$ is equivalent to $T^{a}\equiv\nabla\theta^{a}=0$, $T^{a}$
being the torsion 2-form.

According to the definition given in \S\ref{sec:pc}, the appropriate
\pc\ form for Lagrangian~\eqref{eq:Lec} is
\begin{align}
    \Theta(\L)&\equiv\L + \dV\omega_{ab}\wedge\tfrac{\de\L}{\de\dH\omega_{ab}}
                   \notag  \\
                  &=\L -\tfrac1{2\kappa}\dV\omega_{ab}\wedge\Sigma^{ab},
                   \label{eq:pc}
\end{align}
where $\de\L/\de\dH\omega_{ab}$ stands for $\de L/\de\omega_{ab\nu,\mu}\,
\d s_{\mu\nu}$ and $\d s_{\mu\nu}:=\de_\nu\inn\d s_\mu$. Hence, the Noether
current  associated with a projectable vector field~$\Xi$ is
\begin{align}
E(\L,\Xi) &= -\xi\inn\L
               -\tfrac1{2\kappa} £_\Xi\omega_{ab}\wedge\Sigma^{ab}
               \notag \\
            &\equiv \tfrac1{2\kappa}[(\xi\inn\Omega_{ab})\wedge\Sigma^{ab}
              + \Omega_{ab}\wedge(\xi\inn\Sigma^{ab})
              -£_\Xi\omega_{ab}\wedge\Sigma^{ab}]
               \notag \\
            &\equiv \tfrac1{2\kappa}[(\xi\inn\Omega_{ab})\wedge\Sigma^{ab}
              + \xi^{c}\Omega_{ab}\wedge\Sigma^{ab}\ida_{c}
              -£_\Xi\omega_{ab}\wedge\Sigma^{ab}].
               \label{eq:E}
\end{align}
Now, our configuration bundle~$B$ is $\Sigma_{\rho}\ftimes{M}C$, which is a gauge-natural
bundle. Therefore, every (principal) automorphism $\Phi\in\Aut(\Sigma)$ induces an
automorphism~$\Phi_B$ on~$B$. This holds also infinitesimally, i.e.\ for invariant
(projectable) vector fields defined on~$\Sigma$.  Strictly speaking, an invariant vector field
$\Xi\in\X(\Sigma)$  defines functorially a projectable vector
field~$\Xi_B\in\X(\Sigma_{\rho}\ftimes{M}C)$. Moreover, every $\Spin(1,3)^{e}$-invariant vector
field $\Xi\in\X(\Sigma)$ projects onto an $\SO(1,3)^{e}$-invariant vector field, 
which we denote by the same symbol $\Xi\in\X(\Sigma/\Z_2)$. 
Since the natural projection $\pr\colon\Sigma\to \Sigma/\Z_2$ is a covering map (locally, a
diffeomorphism) of principal fibre bundles, it follows that there is a bijection between
projectable $\SO(1,3)^{e}$-invariant vector fields on $\Sigma/\Z_2$
and projectable $\Spin(1,3)^{e}$-invariant vector fields on $\Sigma$ \cite[\cf\ ][]{fffg96}.
If a spin-frame is given, such a bijection extends to an invariant vector
field bijection between $\Sigma/\Z_2$ and $\SO(M,g)\equiv
\tilde\Lambda(\Sigma)$, and, hence, between $\SO(M,g)$ and
$\Sigma$. Yet, only the Lie derivative of the connection 1-form is needed here, 
so we can simply regard $\Xi_B$ as belonging to 
$\X(C)$. Then, a projectable vector field $\Xi_C\in
\X(C)$ onto a vector field
$\xi\equiv\xi^{\mu}\de_{\mu}\in\X(M)$ reads as
\begin{equation*}
    \Xi_C=\xi^{\mu}\de_{\mu}+\Xi^{a}\ida_{b\mu}\frac{\de}{\de u^{a}\ida_{b\mu}},
\end{equation*}
where
\begin{equation*}
    \Xi^{a}\ida_{b\mu}:=-(\de_{\mu}\xi^{\nu}u^{a}\ida_{b\nu} 
    + u^{a}\ida_{c\mu}\Xi^{c}\ida_{b} - u^{c}\ida_{b\mu}\Xi^{a}\ida_{c}
    +\de_{\mu}\Xi^{a}\ida_{b}),
\end{equation*}
$\Xi\equiv\xi^{\mu}(x)\de_{\mu}+\Xi^{a}\ida_{b}(x)\alpha_{a}\ida^{b}$ 
being the corresponding projectable vector field 
on~$\Sigma/\Z_2$
and $(u^{a}\ida_{b\mu})$ local  fibre coordinates on~$C$.
The vector fields
$\alpha_{a}\ida^{b}$ are local right $\SO(1,3)^{e}$-invariant
vector fields on~$\Sigma/\Z_2$, which in a suitable chart
$(x^\mu, u_a\ida^b)$ read as
\begin{equation*}
\alpha_a\ida^b \equiv\frac12(\rho_a\ida^b -\eta^{bc}\eta_{ad}\rho_c\ida^d),
\end{equation*}
$\eta$ denoting the Minkowski metric and $\rho_a\ida^b:=u_c\ida^b\de/\de u_c\ida^a$.
Therefore, the Lie derivative  of~$u^{a}\ida_{b\mu}=\omega^{a}\ida_{b\mu}(x)$ is just
\begin{equation*}
    £_{\Xi}\omega^{a}\ida_{b\mu}
      = \xi^{\nu}\de_{\nu}\omega^{a}\ida_{b\mu}
        +\de_{\mu}\xi^{\nu}\omega^{a}\ida_{b\nu} 
        + \omega^{a}\ida_{c\mu}\Xi^{c}\ida_{b} 
        - \omega^{c}\ida_{b\mu}\Xi^{a}\ida_{c}
        +\de_{\mu}\Xi^{a}\ida_{b},
\end{equation*}
which can be readily recast in Cartan formalism as
\begin{equation}
    £_{\Xi}\omega^{a}\ida_{b}=\xi\inn\Omega^{a}\ida_{b}+\nabla\VXi^{a}\ida_{b},
    \label{eq:lieconn}
\end{equation}
$\VXi^{a}\ida_{b}:=\Xi ^{a}\ida_{b}+\omega^{a}\ida_{b\mu}\xi^{\mu}$ 
being the vertical part of~$\Xi$.
On substituting~\eqref{eq:lieconn} into~\eqref{eq:E}, we finally get
\begin{align}
	E(\L,\Xi) & = \tfrac1{2\kappa}
	              (\xi^{c}\Omega_{ab}\wedge\Sigma^{ab}\ida_{c}
	               -\nabla\VXi_{ab}\wedge\Sigma^{ab})
	\notag  \\
	 & = \tfrac1{2\kappa}
	              [\xi^{c}\Omega_{ab}\wedge\Sigma^{ab}\ida_{c}
	               +\VXi_{ab}\nabla\Sigma^{ab}
	               -\dH(\VXi_{ab}\Sigma^{ab})].
	\label{eq:E2}
\end{align}
Now, by virtue of equations of motion~\eqref{eq:moto1} 
and~\eqref{eq:moto2},
\begin{equation}
    U(\L,\Xi) := -\frac1{2\kappa}\VXi_{ab}\Sigma^{ab}
    \label{eq:sp}
\end{equation}
is recognized to be the \emphbf{superpotential} associated with 
Lagrangian~\eqref{eq:Lec}. This superpotential, which was derived in a \emph{completely}
gauge-natural context and---to the best of our knowledge---appears here for the first time,
represents the most general superpotential possible in this metric-affine
formulation of gravity (modulo, of course, closed 2-forms).

Note that in the case of the Kosmann lift \cite[]{fffg96} we have
\begin{equation}
    \VXi_{ab} = (\VxiK)_{ab} \equiv - \cd_{[a}\xi_{b]},
    \label{eq:Klift}
\end{equation}
which, substituted in~\eqref{eq:sp}, gives
\begin{equation}
    U(\L,\xiK) = \frac1{2\kappa}\cd_{a}\xi_{b}\Sigma^{ab},
    \label{eq:spK}
\end{equation}
i.e.\ \emph{half} of the well-known \cite{komar59} potential, in accordance with
the result found by \cite{ffm94} in a purely  natural context. This is also the
lift implicitly used by \cite{gmff} in the 2-spinor formalism.

Let now $\sigma_a\ida^{AA'}$ denote the Infeld-van der Waerden symbols, which
express the isomorphism between $\Re[S(M)\otimes\bar S(M)]$ and $TM$ in the orthonormal basis 
induced by the spin-frame chosen \cite[\cf\ ][]{pr1,wald}, and consider the following lift:
\begin{equation}
    \xi^{\mu}=e_{a}\ida^{\mu}\sigma^{a}\ida_{AA'}\l^{A}\bl^{A'},  \qquad
    \VXi_{ab}=(\VxiW)_{ab}:=-4\sigma_{[a}\ida^{AA'}\sigma_{b]}\ida^{BB'}
                              \Re(\bl_{B'}\cd_{BA'}\l_{A}),
    \label{eq:Wlift}
\end{equation}
which will be referred to as the \emphbf{Witten lift}. Then
\begin{equation}
    U(\L,\xiW) = 
      \Re W \equiv -\frac{2}\kappa\Re(\i\bl_{A'}\nabla\l_A\wedge\theta^{AA'}),
    \label{eq:realnw}
\end{equation}
which is the (real) \nw\ 2-form \cite[]{nester81,pr2}. Indeed, we
have\footnote{With the exception of formula \eqref{eq:CWlift} below, we shall
suppress hereafter the Infeld-van der Waerden symbols and adopt the standard
identification $a=AA'$, $b=BB'$, \&c., as is customary in the current literature 
\cite[\cf\ ][]{pr1}.}:
\begin{align}
	\VXi_{ab}\Sigma^{ab} & =-2\bl_{B'}\cd_{BA'}\l_{A}\Sigma^{ab}+\cc
	   \notag  \\
	 & =2\i\liestar{}(\bl_{A'}\cd_{BB'}\l_{A})\Sigma^{ab}+\cc
	   \notag  \\
	 & =2\i\bl_{A'}\cd_{b}\l_{A}\leftsideset{^*}{\Sigma}^{ab}+\cc
	   \notag  \\
	 & =-2\i\bl_{A'}\cd_{b}\l_{A}\theta^{a}\wedge\theta^{b}+\cc
	   \notag  \\
	 & =2\i\bl_{A'}\nabla\l_{A}\wedge\theta^{AA'}+\cc,
	   \label{eq:Wlift2}
\end{align}
where we used the identities \cite[\cf\ ][]{pr1}
\begin{equation*}
    \liestar{A}_{ab}B^{ab}=A_{ab}\liestar{B}^{ab}, \qquad
    \leftsideset{^{**}}{\!A}^{ab}=-A^{ab},\qquad
    \liestar{A}^{ABA'B'}=\i A^{ABB'A'}
\end{equation*}
for any two bivectors $A^{ab}$ and $B^{ab}$.
Inserting~\eqref{eq:Wlift2} into~\eqref{eq:sp} gives 
\eqref{eq:realnw}, as claimed.

If we wish, it is also possible to define a \emphbf{complexified
Witten lift} as
\begin{equation}
    \xi^{\mu}=e_{a}\ida^{\mu}\sigma^{a}\ida_{AA'}\l^{A}\bl^{A'},  \qquad
    \VXi_{ab}=(\VxiCW)_{ab}:=-4\sigma_{[a}\ida^{AA'}\sigma_{b]}\ida^{BB'}
                              \bl_{B'}\cd_{BA'}\l_{A}.
    \label{eq:CWlift}
\end{equation}
Then, the relevant superpotential is
\begin{equation}
    U(\L,\xiCW) = 
      W := -\frac{2\i}\kappa\bl_{A'}\nabla\l_A\wedge\theta^{AA'},
    \label{eq:nw}
\end{equation}
which is the (complex) \nw\ 2-form \cite[]{pr2,mf90}.
From the viewpoint of physical applications (proof of positivity of the
Bondi or ADM mass, quasi-local definitions of momentum and angular 
momentum in general relativity, \&c.), it is immaterial whether one 
uses~\eqref{eq:nw} or its real part~\eqref{eq:realnw}, as its imaginary part 
turns out to be $-1/\kappa\,\dH(\l_{A}\bl_{A'}\theta^{a})$, which 
vanishes upon integration over a closed 2-surface.

\noindent
Note, though, that~\eqref{eq:nw} appears to relate more directly to 
Penrose quasi-local 4-momentum, when suitable identifications are made 
\cite[\cf\ ][p.~432]{pr2}.

\begin{rem}
Note also that---modulo an inessential numerical factor---the Kosmann lift is (the real part 
of) the \emph{dual} of the (complex) Witten lift, in the sense that
\begin{equation*}
(\VxiK)_{ab} = -\frac12\Re[\liestar{}(\VxiCW)_{ab}],
\end{equation*}
as can be easily checked on starting from equations~\eqref{eq:Klift} and~\eqref{eq:CWlift}(2),
whenever, of course, $\xi^a=\l^A\bl^{A'}$.
\end{rem}

\begin{rem}
The theory developed herein is obviously tailored to the coupling with
spinor fields described by the Dirac Lagrangian,
\begin{equation*}
\LD := \left[\frac\i2(\tilde\Psi\gamma^{a}\cd_{a}\Psi
 -\widetilde{\cd_{a}\Psi}\gamma^{a}\Psi) -m\tilde{\Psi}\Psi\right]
\sqrt{g}\,\d s.
\end{equation*}
In the \emph{purely metric} case, the total superpotential
turns out to be
\begin{equation*}
    U(\L+\LD,\Xi) = U(\L,\Xi) + U(\LD,\Xi),
\end{equation*}
where
\begin{align*}
    U(\LD,\Xi) &:= \tfrac\i8\tilde\Psi[(\gamma_{a}\gamma_{b}\gamma_{c}
                   +2\eta_{ac}\gamma_{b})\xi^c]\Psi\,\Sigma^{ab},\\
               &\hphantom{:}\equiv\tfrac{\i\sqrt{2}}{4}\xi_{A}^{A'}
                (\bar\varphi_{A'}\varphi_{B}-\bar\psi_{B}\psi_{A'})\Sigma^{AB}
                      + \cc.
\end{align*}
The reader is referred to \cite{fffg98} and \cite{gmff} for further details and
notation.

\noindent
Conversely, in the present \emph{metric-affine} context, it can be readily shown that, although 
the Dirac Lagrangian {\it does\/} enter the equations of motion (notably, the ``second''
\emph{Einstein-Cartan} equation), it {\it does not\/} contribute to the total
superpotential. From this fact one might mistakenly conclude that the Dirac fields do
not contribute to the total conserved quantities. This conclusion is wrong because,
although the Dirac Lagrangian does not contribute directly to the superpotential, in
order to obtain the corresponding conserved quantities, one needs integrate the
superpotential on a solution, which in turn depends on the Dirac Lagrangian via its
energy-momentum tensor and the second Einstein-Cartan equation.
\end{rem}

\section{First order gravity}
\label{sec:fog}

In the case of \emph{vanishing torsion} ($T^a\equiv0\iff\nabla\Sigma^{ab}\equiv0$),
it is easy to see that Lagrangian~\eqref{eq:Lec} can be split into a total
divergence plus a first order covariant Lagrangian. In many contexts, the
superpotential associated with this Lagrangian proved to give more physically reasonable 
answers than the Hilbert superpotential \cite[\cf\ ][]{gmff}. 

For this reason and the sake of completeness, we now give the derivation of the
aforementioned superpotential in the new geometrical framework outlined in
\S\ref{sec:gf}.

The first order covariant Lagrangian in question is \cite[\cf\ ][]{ff90,ffm94}
\begin{align}
	\hat\L & := -\tfrac1{2\kappa}(\hat\Omega_{ab} - Q_{ac}\wedge Q^c\ida_b)
               \wedge\Sigma^{ab}
	            \label{eq:Lbg}  \\
	       & \hphantom{:}\equiv \L + \tfrac{1}{2\kappa}\dH(Q_{ab}\wedge\Sigma^{ab}),
	            \notag
\end{align}
where $\L$ is given by~\eqref{eq:Lec}, $\hat\Omega_{ab}:=\dH\hat\omega_{ab}
+\hat\omega_{ac}\wedge\hat\omega^c\ida_b$ and
$Q_{ab}:=\omega_{ab}-\hat\omega_{ab}$, $\hat\omega_{ab}$ being a ``background''
(non-dynamical) spin-connection. The corresponding \pc\ form is
\begin{align*}
    \Theta(\hat\L)&=\hat\L  - \tfrac1{2\kappa}(\dV\hat\omega_{ab}\wedge\Sigma^{ab}
                    - \dV\Sigma^{ab}\wedge Q_{ab}) \\
                  &\equiv\Theta(\L) + \tfrac1{2\kappa}[\dH(Q_{ab}\wedge\Sigma^{ab})
                    +\dV Q_{ab}\wedge\Sigma^{ab} + \dV\Sigma^{ab}\wedge Q_{ab}],
\end{align*}
Hence, the Noether current associated with a projectable vector 
field~$\Xi$ is 
\begin{align}
E(\hat\L,\Xi) &= E(\L,\Xi) + \tfrac1{2\kappa}[£_{\Xi}Q_{ab}\wedge\Sigma^{ab}
                  +£_{\Xi}\Sigma^{ab}\wedge Q_{ab}
                  -\xi\inn\dH(Q_{ab}\wedge\Sigma^{ab})]
                 \notag \\
              &= E(\L,\Xi) + \tfrac1{2\kappa}[£_{\Xi}Q_{ab}\wedge\Sigma^{ab}
                  +£_{\Xi}(Q_{ab}\wedge\Sigma^{ab})
                  -£_{\Xi}Q_{ab}\wedge\Sigma^{ab}
                 \notag \\
              &\qquad\qquad\qquad\quad{}-\xi\inn\dH(Q_{ab}\wedge\Sigma^{ab})] 
                 \notag \\ 
              &= E(\L,\Xi) + \tfrac1{2\kappa}
                 [£_{\Xi}(Q_{ab}\wedge\Sigma^{ab})
                  -\xi\inn\dH(Q_{ab}\wedge\Sigma^{ab})],
                 \label{eq:Ebg}
\end{align}
$\xi$ denoting, as usual, the projection of~$\Xi$ onto~$M$. Now,
\begin{align}
    £_{\Xi}(Q_{ab}\wedge\Sigma^{ab})&\equiv £_{\xi}(Q_{ab}\wedge\Sigma^{ab})
    \notag \\
    &=\xi\inn\dH(Q_{ab}\wedge\Sigma^{ab})+\dH[\xi\inn(Q_{ab}\wedge\Sigma^{ab})].
    \label{eq:lieQSigma}
\end{align}
On substituting~\eqref{eq:lieQSigma} into~\eqref{eq:Ebg}, we get
\begin{equation*}
	E(\hat\L,\Xi) = E(\L,\Xi) + \frac1{2\kappa}
                    \dH[\xi\inn(Q_{ab}\wedge\Sigma^{ab})],
\end{equation*}
whence
\begin{equation}
    U(\hat\L,\Xi) := U(\L,\Xi) + \frac1{2\kappa}\xi\inn(Q_{ab}\wedge\Sigma^{ab})
\end{equation}
is recognized to be the superpotential associated with 
Lagrangian~\eqref{eq:Lbg}.

\begin{rem}
Note that, contrary to what happens in the purely natural context, no additional
conditions need be imposed on the vector field~$\Xi$ here.
\end{rem}

\section*{Discussion}

This paper stresses the important role the 
theory of gauge-natural bundles plays in a significant 
issue of mathematical physics such as the definition of the 
gravitational energy and, more generally, of conserved quantities 
associated with the gravitational field, especially when coupled to
spinor fields.

Besides providing a new gravitational superpotential in a 
gauge-natural context, the paper sheds some new light on the 
definition of the \nw\ 2-form and gives it an interpretation as a 
further, genuine gravitational superpotential. 

Moreover, this paper shows that it is crucial in this context \emph{not} to 
regard the metric as the fundamental gravitational field. Indeed, 
when considering the interaction between gravity 
and spinors one is forced to give up a purely natural formalism and
instead consider a gauge-natural formalism in which one chooses a 
spin-tetrad (together with a spin-connection, in a metric-affine 
formulation) as one's fundamental variable.

A parallel and analogous method of investigation is possible when dealing, 
in a gauge-natural context, with Legendre and dual Legendre transforms, 
for which the reader is referred to the recent and fundamental papers 
by \cite{raiteri96} and \cite{raiteri00}.

\section*{Acknowledgements}

P. M. acknowledges an EPSRC research studentship and a Faculty Research
Studentship from the University of Southampton.

\bibliographystyle{pmbib}

\begin{thebibliography}{}

\bibitem[Dubois-Violette \& Madore, 1987]{dubois-madore87}
Dubois-Violette, M. \& Madore, J. (\oldstylenums{1987}).
\newblock `Conservation laws and integrability conditions for gravitational and
  Yang-Mills field equations'.
\newblock \emph{Comm. Math. Phys.} \textbf{108}, 213--23.

\bibitem[Fatibene \emph{et~al.{}}, 1996]{fffg96}
Fatibene, L., Ferraris, M., Francaviglia, M. \& Godina, M.
  (\oldstylenums{1996}).
\newblock `A geometric definition of Lie derivative for spinor fields'.
\newblock In J.~Jany{\v s}ka, I.~Kol{\'a}{\v r} \& J.~Slov{\'a}k (Eds.),
  \emph{Proc. 6th International Conference on Differential Geometry and its
  Applications {\textup{(Brno, {\anno{1995}})}}}, pp.\ 549--58. Brno: Masaryk
  University.

\bibitem[Fatibene \emph{et~al.{}}, 1998]{fffg98}
Fatibene, L., Ferraris, M., Francaviglia, M. \& Godina, M.
  (\oldstylenums{1998}).
\newblock `Gauge formalism for general relativity and fermionic matter'.
\newblock \emph{Gen. Rel. Grav.} \textbf{30}(9), 1371--89.

\bibitem[Ferraris \& Francaviglia, 1990]{ff90}
Ferraris, M. \& Francaviglia, M. (\oldstylenums{1990}).
\newblock `Covariant first-order Lagrangians, energy-density and
  superpotentials in general relativity'.
\newblock \emph{Gen. Rel. Grav.} \textbf{22}(9), 965--85.

\bibitem[Ferraris \emph{et~al.{}}, 1994]{ffm94}
Ferraris, M., Francaviglia, M. \& Mottini, M. (\oldstylenums{1994}).
\newblock `Conserved quantities of the gravitational field in tetrad notation'.
\newblock \emph{Rend. Mat. Appl.~(7)}~\textbf{14}, 457--81.

\bibitem[Ferraris \emph{et~al.{}}, 2000]{raiteri00}
Ferraris, M., Francaviglia, M. \& Raiteri, M. (\oldstylenums{2000}).
\newblock `Dual Lagrangian field theories'.
\newblock \emph{J.~Math. Phys.} \textbf{41}(4), 1889--915.

\bibitem[Ferraris \emph{et~al.{}}, 1986]{ffr86}
Ferraris, M., Francaviglia, M. \& Robutti, O. (\oldstylenums{1986}).
\newblock `Energy and superpotentials in gravitational field theories'.
\newblock In M.~Modugno (Ed.), \emph{Atti del 6{$^{\text o}$} Convegno
  Nazionale di Relativit{\`a} Generale e Fisica della Gravitazione}, pp.\
  137--50. Bologna: Pitagora Editrice.

\bibitem[Giachetta \emph{et~al.{}}, 1997]{gms}
Giachetta, G., Mangiarotti, L. \& Sardanashvily, G. (\oldstylenums{1997}).
\newblock \emph{New Lagrangian and Hamiltonian methods in field theory}.
\newblock Singapore: World Scientific.

\bibitem[Godina \emph{et~al.{}}, 2000]{gmff}
Godina, M., Matteucci, P., Fatibene, L. \& Francaviglia, M.
  (\oldstylenums{2000}).
\newblock `Two-spinor formulation of first order gravity coupled to Dirac
  fields'.
\newblock \emph{Gen. Rel. Grav.} \textbf{32}(1), 145--59.

\bibitem[Gotay \emph{et~al.{}}, 1998]{gimmsy1}
Gotay, M.~J., Isenberg, J. \& Marsden, J.~E. (\oldstylenums{1998}).
\newblock \emph{{\textup{`Momentum maps and classical relativistic fields.
  Part~I: Covariant field theory'.}}}
\newblock E-print: physics{@}xxx.lanl.gov ({\emph{physics/9801019}}).

\bibitem[Kijowski, 1973]{kijowski73}
Kijowski, J. (\oldstylenums{1973}).
\newblock `A finite-dimensional canonical formalism in the classical field
  theory'.
\newblock \emph{Comm. Math. Phys.}~\textbf{30}, 99--128.

\bibitem[Kijowski, 1978]{kijowski78}
Kijowski, J. (\oldstylenums{1978}).
\newblock `On a new variational principle in general relativity and the energy
  of the gravitational field'.
\newblock \emph{Gen. Rel. Grav.} \textbf{9}(10), 857--77.

\bibitem[Kijowski \& Szczyrba, 1975]{ks75}
Kijowski, J. \& Szczyrba, W. (\oldstylenums{1975}).
\newblock `Multisymplectic manifolds and the geometrical construction of the
  Poisson brackets in the classical field theory'.
\newblock In J.-M. Souriau (Ed.), \emph{G{\'e}om{\'e}trie Symplectique et
  Physique Math{\'e}matique {\textup{(Colloq. Internat. C.N.R.S.,
  Aix-en-Provence, {\anno{1974}})}}}, pp.\ 347--79. Paris: C.N.R.S.

\bibitem[Kijowski \& Szczyrba, 1976]{ks76}
Kijowski, J. \& Szczyrba, W. (\oldstylenums{1976}).
\newblock `A canonical structure for classical field theories'.
\newblock \emph{Comm. Math. Phys.}~\textbf{46}, 183--206.

\bibitem[Kijowski \& Tulczyjew, 1979]{kt79}
Kijowski, J. \& Tulczyjew, W.~M. (\oldstylenums{1979}).
\newblock \emph{A symplectic framework for field theories}.
\newblock Lecture Notes in Physics \textbf{107}. Berlin: Springer-Verlag.

\bibitem[Kol{\'a}{\v r} \emph{et~al.{}}, 1993]{kms}
Kol{\'a}{\v r}, I., Michor, P.~W. \& Slov{\'a}k, J. (\oldstylenums{1993}).
\newblock \emph{Natural operations in differential geometry}.
\newblock Berlin: Springer-Verlag.

\bibitem[Komar, 1959]{komar59}
Komar, A. (\oldstylenums{1959}).
\newblock `Covariant conservation laws in general relativity'.
\newblock \emph{Phys. Rev.} \textbf{113}, 934--6.

\bibitem[Krupka, 1971]{krupka71}
Krupka, D. (\oldstylenums{1971}).
\newblock `Lagrange theory in fibered manifolds'.
\newblock \emph{Rep. Mathematical Phys.}~\textbf{2}, 121--33.

\bibitem[Mason \& Frauendiener, 1990]{mf90}
Mason, L.~J. \& Frauendiener, J. (\oldstylenums{1990}).
\newblock `The Sparling 3-form, Ashtekar variables and quasi-local mass'.
\newblock In T.~N. Bailey \& R.~J. Baston (Eds.), \emph{Twistors in mathematics
  and physics}, London Math.\ Soc.\ Lecture Notes \textbf{156}, pp.\ 189--217.
  Cambridge: Cambridge University Press.

\bibitem[Nester, 1981]{nester81}
Nester, J.~M. (\oldstylenums{1981}).
\newblock `A new gravitational energy expression with a simple positivity
  proof'.
\newblock \emph{Phys. Lett.} \textbf{83A}, 241--2.

\bibitem[Penrose, 1982]{penrose82}
Penrose, R. (\oldstylenums{1982}).
\newblock `Quasi-local mass and angular momentum in general relativity'.
\newblock \emph{Proc. Roy. Soc. London Ser. A} \textbf{381}, 53--62.

\bibitem[Penrose \& Rindler, 1984]{pr1}
Penrose, R. \& Rindler, W. (\oldstylenums{1984}).
\newblock \emph{Spinors and space-time. Vol.~1: Two-spinor calculus and
  relativistic fields}.
\newblock Cambridge: Cambridge University Press.

\bibitem[Penrose \& Rindler, 1986]{pr2}
Penrose, R. \& Rindler, W. (\oldstylenums{1986}).
\newblock \emph{Spinors and space-time. Vol.~2: Spinor and twistor methods in
  space-time geometry}.
\newblock Cambridge: Cambridge University Press.

\bibitem[Raiteri \emph{et~al.{}}, 1996]{raiteri96}
Raiteri, M., Ferraris, M. \& Francaviglia, M. (\oldstylenums{1996}).
\newblock `General relativity as a gauge theory of orthogonal groups in three
  dimensions'.
\newblock In P.~Pronin \& G.~Sardanashvily (Eds.), \emph{Gravity, particles and
  space-time}, pp.\ 81--98. Singapore: World Scientific.

\bibitem[Trautman, 1967]{trautman67}
Trautman, A. (\oldstylenums{1967}).
\newblock `Noether equations and conservation laws'.
\newblock \emph{Comm. Math. Phys.}~\textbf{6}, 248--61.

\bibitem[Wald, 1984]{wald}
Wald, R.~M. (\oldstylenums{1984}).
\newblock \emph{General relativity}.
\newblock Chicago: The University of Chicago Press.

\end{thebibliography}

\providecommand{\SortNoop}[1]{}

\end{document}